# Synthesis of titanium-oxide nanoparticles with size and stoichiometry control

**Rickard Gunnarsson, Ulf Helmersson** and **Iris Pilch**

Department of Physics, Chemistry and Biology (IFM),
Linköping University, SE-581 83 Linköping, Sweden

E-mail: ricgu@ifm.liu.se, ulfhe@ifm.liu.se, iripi@ifm.liu.se

## Abstract

Ti-O nanoparticles have been synthesized via hollow cathode sputtering in an Ar-$O_2$ atmosphere using high power pulsing. It is shown that the stoichiometry and the size of the nanoparticles can be varied independently, the former through controlling the $O_2$ gas flow and the latter by the independent biasing of two separate anodes in the growth zone. Nanoparticles with diameters in the range of 25 to 75 nm, and with different Ti-O compositions and crystalline phases, have been synthesized.

## Background

$TiO_2$ nanoparticles (NPs) are used in a wide range of applications, and one of their well-known properties is photocatalytic activity. Since the physical properties of NPs depend on the NP characteristics such as size and composition, it is essential to have a synthesis process in which these properties are controllable. Thus, much attention has been focused on the nanometer-scale structure and chemical composition of $TiO_2$ [1]. One method that has been shown to be able to control the stoichiometry of Ti-O thin films is reactive sputtering [2]. A problem that can arise during reactive sputtering is oxidization of the cathode surface, which affects the process conditions and decreases the flux of Ti from the cathode substantially [3].

The situation when a cathode becomes oxidized is referred to as poisoning. When the cathode is poisoned, the metal sputter rate typically decreases and the surface conductivity decreases, which increases the tendency for formation of arcs [4]. The degree of poisoning depends on the flow of reactive gas to the surface, the reaction probability with the cathode, and the removal rate of the reacted layer by sputtering. Therefore, the process conditions are often not the same for the same reactive gas flow depending on whether the gas flow is increased or decreased. This hysteresis in the process can be observed in several different process parameters, *e.g.* the voltage and the reactive gas partial pressure [3]. For thin film synthesis of $TiO_2$ using sputtering, it is often found that an optimum, with respect to the deposition rate of the film, is obtained by operation in the hysteresis regime. However, in this regime the process is usually unstable and a fast feedback control of the $O_2$ flow is required to maintain stable operation [4]. Under certain conditions, the instability in the desirable operation region for discharge operation can be avoided by for example using small cathode surface areas to increase the removal rate of the reacted layer [5], by using high power impulse magnetron sputtering, HiPIMS [6], or by techniques limiting the flow of reactive gas to the sputtering source surface, *e.g.*, through baffles and a separate inert gas flow to the area around the source [7]. A different approach from using magnetron sputtering for reactive synthesis of thin films is to use a hollow cathode, which has been shown to offer a high growth rate of $Al_2O_3$ thin films [8]. Cathode poisoning is here reduced as the result of the high flow of inert gas through the hollow cathode which suppresses the reactive gas that is fed into the system closer to the substrate from entering into the cathode [9] [8].

NPs synthesis, using reactive magnetron sputtering, faces similar problems with poisoning of the cathode [10][11][12] [13]. For nanoparticle nucleation and growth in the gas phase, a high density of sputtered material is required. The highest density is found close to the cathode, which poses a problem for synthesizing fully stoichiometric oxide particles without poisoning the cathode surface which gives a risk of reduced productivity, analogous to the lower deposition rate of thin films using poisoned targets. Methods for increasing the efficiency have been investigated. An example is the use of pulsed direct current power operation where a substantial increase in productivity could be reported in the production of stoichiometric $TiO_2$ NPs [14].

A different approach to increase the growth speed of the NPs is by utilizing a high power pulsed hollow cathode, which results in a high degree of ionization of the sputtered material [15]. This in turn result in a more efficient trapping of ions than neutrals on nucleated NPs that are charged negative in a plasma [16]. The size has also been shown to be controllable either by altering the amplitude, length, and repetition frequency of the pulses, or by varying the discharge geometry outside the hollow cathode [17] [16].

In the present work, the method utilizing high power pulses for sputtering of a hollow cathode is used to grow Ti-O NPs. It is shown that an independent control over the stoichiometry and the size of the NPs can be achieved.

## Methods

The experiments were conducted in a high vacuum system with a base pressure in the low $10^{-6}$ Torr ($10^{-4}$ Pa) range. Before experiments the base pressure was further reduced to the low $10^{-7}$ Torr ($10^{-5}$ Pa) range by getter pumping, i.e. sputtering Ti from a planar magnetron source. A sketch of the setup is shown in Fig. 1(a). The chamber is cylindrical with a diameter of 200 mm and a height of 450 mm. A hollow cathode is mounted in a water-cooled polyoxymethylene holder on the lid of the chamber. The cathode is made of a Ti cylinder with an inner diameter of 5 mm and a length of 55 mm. A grounded anode ring (diameter 30 mm) was placed 45 mm below the hollow cathode. The anode ring and the growth region of NPs are enclosed by a stainless steel mesh with diameter of 90 mm and length of 120 mm. This mesh is kept at a negative bias potential and has a 20 mm hole in the bottom for the NPs to pass through. The setup allows operating at average powers of up to 100 W without a risk of melting the cathode holder. The power to the cathode was supplied by an MDX 1K DC generator connected to a homemade pulsing unit that supplies square voltage pulses with an amplitude of $U_D$. The DC power supply was set in current regulation mode and held at 0.52 A, the pulse width was set to 80 μs and the frequency to 1.5 kHz. This resulted in $U_D$ around 290 V and peak discharge currents of the pulses around 10 A. A bias in the range of -1 V to -50 V was applied to the mesh by connecting a unipolar DC power supply via an active clamp circuit. The primary objective of the negative mesh bias was to steer negatively charged NPs to the exit opening above the substrate, see Fig. 1. However, during the pulse-on time of the discharge, the active clamp also allowed current to flow to the mesh. At all applied biases the mesh was positive with respect to the hollow cathode and thereby could act as an extra anode, but it only drew current in the range of -1 to -23 V. By varying the mesh bias, a variable fraction of the current could thus in this range be diverted from going to the anode ring, downwards in Fig. 1, to flow more in the radial directions towards the mesh.

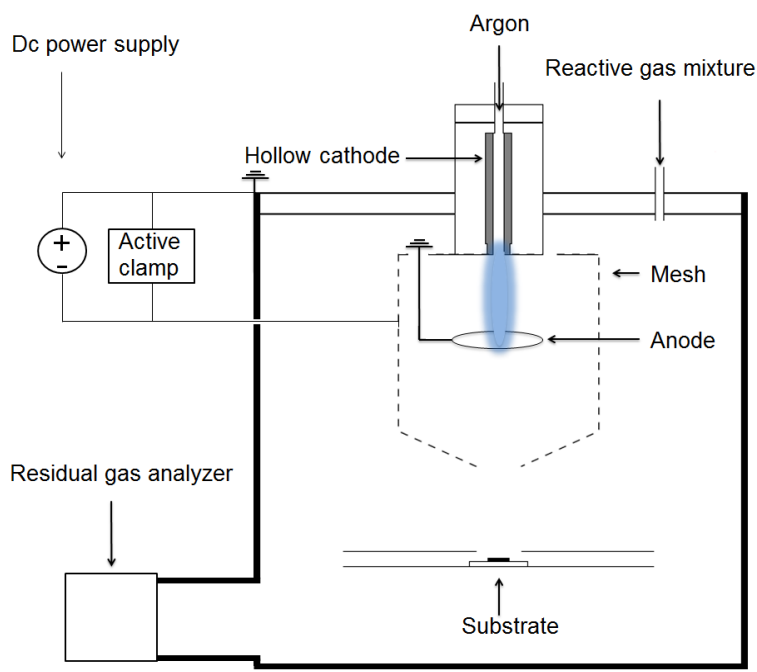

Fig.1. A sketch of the experimental setup (not drawn to scale). The setup consists of a hollow cathode that is driven by high power pulses (1.5 kHz, 80 μs) as the sputter cathode, a grounded anode ring, and a mesh that can be biased. The nanoparticles are steered by the mesh towards the opening at the bottom and deposited on a substrate that is biased positive in order to attract them.

Two different gas inlets were used. The main process gas, Ar, was fed through the hollow cathode at a flow rate of 90 sccm. The $O_2$, which was diluted with 95 mol% Ar to increase flow accuracy, was feed through an inlet in the chamber lid. In this way the $O_2$ flow was varied between 0 and 0.175 sccm (i.e., 3.5 sccm of the Ar-$O_2$ gas admixture), with experiments always starting at low reactive gas flows. The total pressure was set to 0.83 Torr (111 Pa) and automatically regulated by a throttle valve in front of the pump. The process gas was analyzed with a Spectra microvision differentially pumped residual gas analyzer (RGA). The $O_2$ partial pressure in the process chamber was determined by assuming the $O_2$ pressure to be proportional with that in the RGA using the $^{38}$Ar peak as reference value.

The effect of cathode poisoning was investigated by increasing the $O_2$ flow in steps from 0 to 0.175 sccm, and then decreasing it in the same steps. The discharge pulse voltage $U_D$ and the partial pressure of $O_2$ were measured as functions of the $O_2$ flow. After each change in the $O_2$ flow a waiting time of two minutes was used before measuring the discharge voltage and the partial pressure.

The synthesized NPs were collected on substrates (10 mm by 10 mm Au-coated Si), which were positively biased to attract NPs. The positive substrate bias value was kept constant at 10 V for samples used for determining the size distribution. Substrates deposited for X-ray diffraction (XRD) analyses were kept at a slightly higher voltage of up to 15 V in order to attract more NPs and thus improve the signal-to-noise ratio of the XRD analysis. Higher substrate biases could not be used since the substrate then, occasionally, acted as an additional anode and a plasma column formed from the hollow cathode to the substrate.

The NPs were analyzed using a LEO 1550 Gemini scanning electron microscope (SEM) equipped with an energy dispersive X-ray (EDX) detector. The crystal structure of the NPs was determined using XRD operated in a grazing incidence geometry. The NP size distributions were calculated from SEM images, which were taken in a line from one edge of the substrate to another, by using an image analysis program written in Matlab.

## Results

### Characterization of the Discharge Properties

To examine the process characteristics with respect to cathode poisoning and hysteretic behavior the discharge voltage $U_D$, and the $O_2$ partial pressure, were measured for increasing and decreasing $O_2$ gas flows. The $O_2$ gas flow was increased step-wise to a maximum flow rate of 0.175 sccm and then decreased in the same steps. One example each of the measured oxygen partial pressure and the discharge voltage is shown in Fig. 2. A small hysteresis is observed in both curves. However, it was observed that the width of the hysteresis depends on the waiting time after changing the flow. This indicates that the occurrence of a hysteresis is because the equilibrium of the process has not been reached rather than due to an unstable operation as often found in reactive magnetron sputtering. In order to obtain reproducible results without using excessive waiting times we have always taken data, and produced NPs, on the branch with increasing $O_2$ flow and with a waiting time of 2 minutes unless otherwise stated. Repeated measurement of the partial pressure behavior showed close to identical results to the curves shown in Fig. 2. With increasing gas flow (filled circles), the $O_2$ partial pressure was below the detection limit of the RGA for $O_2$ flows up to 0.025 sccm. For higher gas flows the $O_2$ partial pressure becomes detectable and, above 0.100 sccm, a linear increase is found. The linear increase rate corresponds to the increase rate of the $O_2$ partial pressure as a function of $O_2$ gas flow without the plasma discharge turned on. This indicates that the process of gettering oxygen on the sputtered Ti is saturated in the linear range. In a separate experiment (not shown), the $O_2$ partial pressure was measured for higher $O_2$ gas flows up to 0.400 sccm with the same linear trend. For even higher $O_2$ gas flows, the discharge plasma changed color from blue to purple and arcing became a problem.

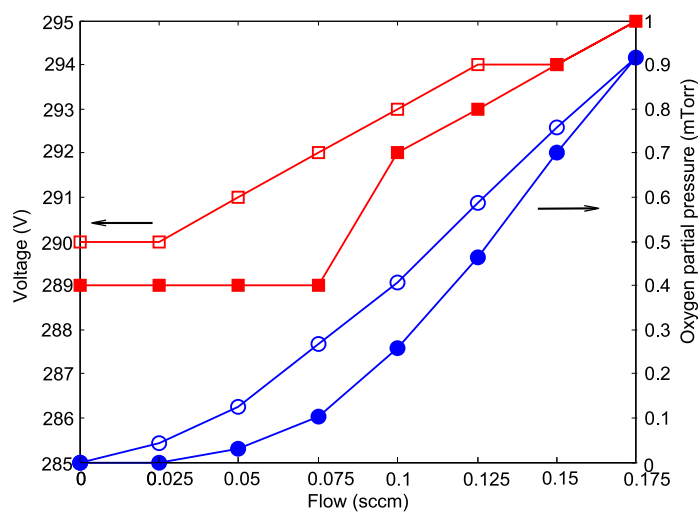

Fig. 2. Examples of the oxygen partial pressure (circles) and the discharge voltage $U_D$ (squares) as functions of the oxygen flow. Filled (open) symbols denote increasing (decreasing) gas flow. The hysteresis that is observed, for both the oxygen partial pressure and the discharge voltage, depends on the waiting time after changing flow, here 2 minutes. The curves for the $O_2$ partial pressure are highly reproducible, while the curves for $U_D$ vary by typically 1- 2 V around the typical curves shown here.

The discharge voltage $U_D$ followed a similar trend as the partial pressure, both for increasing and decreasing $O_2$ gas flows, but was much less reproducible. The curves for $U_D$ varied by typically 1- 2 V around the typical curves shown in Fig. 2, *i.e.,* the variations were of the same order as the observed hysteresis. The main conclusion that holds in spite of these variations is that, above 0.100 sccm, there is a monotonic increase in voltage with oxygen gas flow and with no tendency of saturation. Taken together with the features of the $O_2$ partial pressure, these results indicate that oxidation of the cathode sets in somewhere between 0.025 and 0.100 sccm. Although the cathode is influenced by poisoning, the cathode is far from fully poisoned in the oxygen flow range in which we synthesize NPs (up to 0.15 sccm) since arcing sets in at flows above 0.4 sccm and no sign of saturation of the increase at $U_D$ at 0.175 sccm was found.

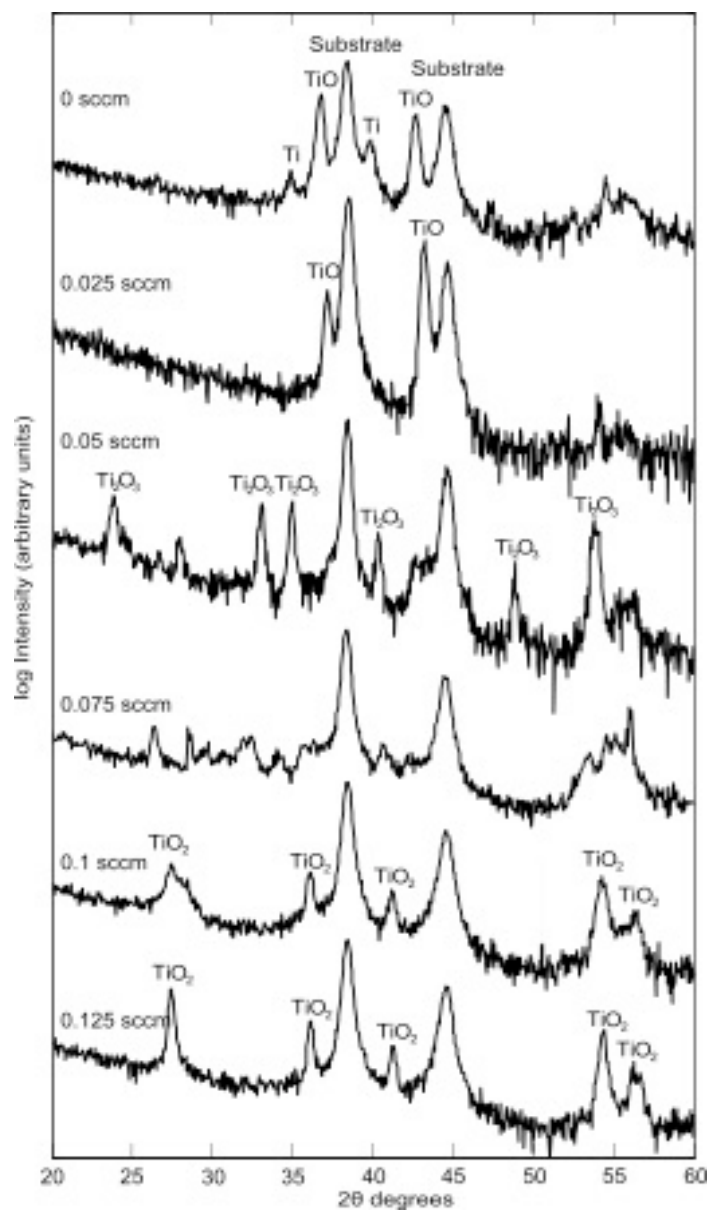

Fig. 3. XRD measurements of the nanoparticles synthesized at different $O_2$ flows. It is found that with higher $O_2$ flow the oxygen-rich phases take over. The diffractograms show also the peaks originated from the gold substrate.

## Size and Stoichiometry Controlled Synthesis of Titanium Oxide Nanoparticles

In order to investigate the variation in stoichiometry, NPs were synthesized using different reactive gas flow rates between 0 sccm (*i.e.*, no inlet of oxygen) and 0.125 sccm. The NPs were then characterized by XRD and EDX. The resulting x-ray diffractograms for one such series are shown in Fig. 3. At 0 sccm, a mixture of Ti and TiO phase was observed, however these peaks were not reproducible and repeated experiments were hard to evaluate by XRD.

For increasing O$_2$ flows, the phase changed from pure TiO at 0.025 sccm, over to Ti$_2$O$_3$ at 0.050 sccm, and finally to TiO$_2$ (rutile) at 0.100 and 0.125 sccm. The stoichiometry observed at a flow rate of 0.075 sccm is less clear and the peaks showed sometimes mixed compositions, *e.g.* Ti$_2$O$_3$ and Ti$_3$O$_5$, or only the substrate peaks could be assigned, as in Fig. 3. We propose that this flow lies right at a transition region between flows that give pure compositions, and that the stoichiometry of the NPs therefore here is very sensitive to drifts and variations in the process. At this transition there is also a change in color of the deposited NPs: at a flow of ≤ 0.075 sccm, the deposited NPs on the substrate had a dark color, and above 0.075 sccm the white color of the fully oxidized TiO$_2$ phase. The EDX measurements, not shown, revealed the same trend that the oxygen content increased with increasing reactive gas flow. A significant amount of oxygen was measured in all samples including the ones synthesized using no O$_2$ flow. Possible reasons for this are residual oxygen in the chamber or exposure to air before the analysis. The size of the NPs was relatively unaffected by the O$_2$ flow. The average diameters were ranging between 57 and 64 nm with the mesh bias set to -50V.

In earlier work with a similar setup, but with a Cu cathode and in a pure Ar gas, two methods to control the NP size were demonstrated by Pilch *et al* [17] [16]. The first method is to change the density of sputtered growth material that is ejected from the hollow cathode orifice into the growth zone, through varying the pulse parameters. Stronger pulses, and/or higher pulse frequencies, generally gave larger NPs. This method was found to be impractical in our reactive discharge because, at low discharge powers, the productivity decreased and the reproducibility became poor due to cathode oxidation. The other method is built on the fact that nanoparticles grow faster by ion collection in plasmas with a higher electron temperature. They then acquire a more negative potential and attract the ions more efficiently. The volume with heated electrons was in Ref. [16] varied by changing the distance between the cathode and the anode ring, see Fig 1. Here, we instead vary the negative bias applied to the mesh cage. When the mesh bias was varied, the discharge was operated at the same peak pulse current, and the O$_2$ partial pressure was found to be uninfluenced. This allows us to assume – in a first approximation – that the mesh bias has little influence on the outflow of sputtered material from the hollow cathode orifice, and that the NP size should not change for this reason. However, the current path through the plasma changed. For small negative values, a large fraction of the current goes through mesh (for -1 V about ¾ of the current passes through the mesh) while at high negative mesh biases very little go through the mesh (at -23 V only a few percent goes through the mesh). According to model calculations [15]. the growth material is carried downward with the outflow from the hollow cathode. When the mesh is used as a separate anode this allows us, in the range -1 to -23 V to control the fraction of the current that heats the plasma, around the growth material, in the volume between the hollow cathode and the anode ring (see Fig. 1).

The mean particle size and the standard deviations are shown in Fig. 4 for different mesh biases with NPs synthesized at O$_2$ gas flow rates of 0.050 sccm [Fig. 4(a)] and 0.150 sccm [Fig. 4(b)]. In total, six experiments were performed showing the same general trend of decreasing size with decreasing negative mesh bias, *i.e.*, the absolute value of the mesh bias decreased. At -50 V, the NPs have a size of about 70 nm with a wide spread of the size distribution. The size of the NPs decreases nearly linear to an average size of 25 nm at -1 V. Note that the sizes of the NPs are similar for the two $O_2$ flows. The standard deviation is typically 10 % of the absolute value, except at -50 V and -1 V. A reason why the size distribution is broader at -50 V is that some NPs tend to be coagulated and were detected as single NPs, see Fig. 5(a). When the absolute value of the negative mesh bias was decreased, the fraction of coagulated NPs decreased and the NPs had a spherical shape as shown in Fig. 5(b) for a mesh bias of -1 V. It has to be pointed out that the process tends to change in time until reaching equilibrium. Similar behavior has previously been reported as a drift in the deposition rate [12]. The initial drift in our process was observed to lead to a decrease in coagulated NPs in time and a slight decrease in size, which is reflected as spread of the average particle diameter in Fig. 4. The decrease in size due to a drift of the process is less than the observed change in size by changing the mesh bias. Hence, it can be concluded that the mesh voltage and not a drift in the process parameters is the cause for the change in diameter.

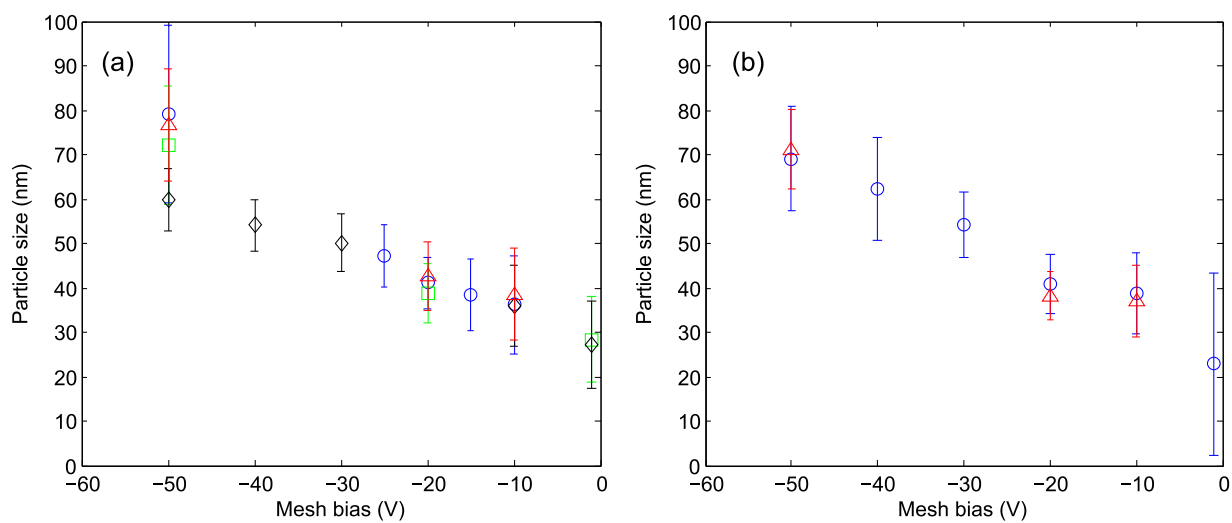

Fig. 4. The mean diameter and the standard deviation of the nanoparticle size distribution as function of mesh bias for $O_2$ gas flow of (a) 0.050 and (b) 0.150 sccm. A clear trend of smaller NP size with less negative mesh bias is found. The different symbols indicate experiments performed at different occasions but nominally identical conditions.

In the two experimental series described above, the experiments with varying $O_2$ flow were analyzed with focus on the stoichiometry, and the experiments with variable mesh bias were analyzed with focus on the size. A third set of experiments addresses the questions how much the $O_2$ flow influences the size, and how much the mesh bias influences the stoichiometry. Here both the $O_2$ flow and the mesh bias were varied, and both the size and the stoichiometry were measured. To be able to do this, two samples were synthesized under identical conditions but with different deposition times. A short deposition time of two minutes was used in order to get well-separated particles to measure the size from SEM images, and a longer deposition time of 20 minutes was used to accurately determine the crystal phase of the NPs with XRD. The mean size and the standard deviation are shown in Fig. 6 (a) as functions of the mesh bias. Five mesh bias voltages were used, and four different $O_2$ flow rates. Independent of $O_2$ flow, the size variation with mesh bias agrees with Fig. 4 and, independent of mesh bias, the stoichiometry changed from TiO at 0.025 sccm to $TiO_2$ at 0.150 sccm as shown in Fig 6. (b). Both measurement points at 0.150 sccm revealed stoichiometric $TiO_2$. For TiO XRD data was available for three points (-50, -20 and -1 V) and only for the mesh bias of -1 V the signal was weak due to a low deposition of NPs but still the measurement indicated that the NPs were TiO. For 0.050 sccm, a variation of the observed crystalline phases was found: for one sample (red triangle) two crystalline phases, $Ti_2O_3$ and $Ti_3O_5$, were measured whereas for another (green square) only $Ti_2O_3$ was found. Although different crystalline phases were synthesized, the phase remained the same during the experiment when the mesh bias was changed, which shows that the crystalline phase indeed can be retained during operation and NPs with the same crystalline phase but different sizes can be synthesized. In summary, within the experimental accuracy the size is determined only by the mesh bias, and the stoichiometry only by the oxygen flow.

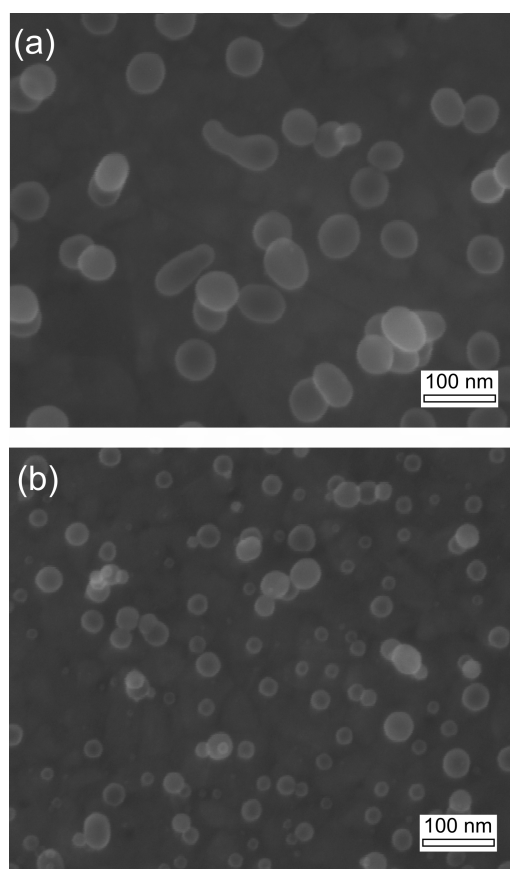

Fig. 5 SEM images of nanoparticles produced at $O_2$ flow of 0.050 sccm at -50 V mesh bias (a) and -1 V mesh bias (b). The nanoparticle size is smaller for less negative mesh bias.

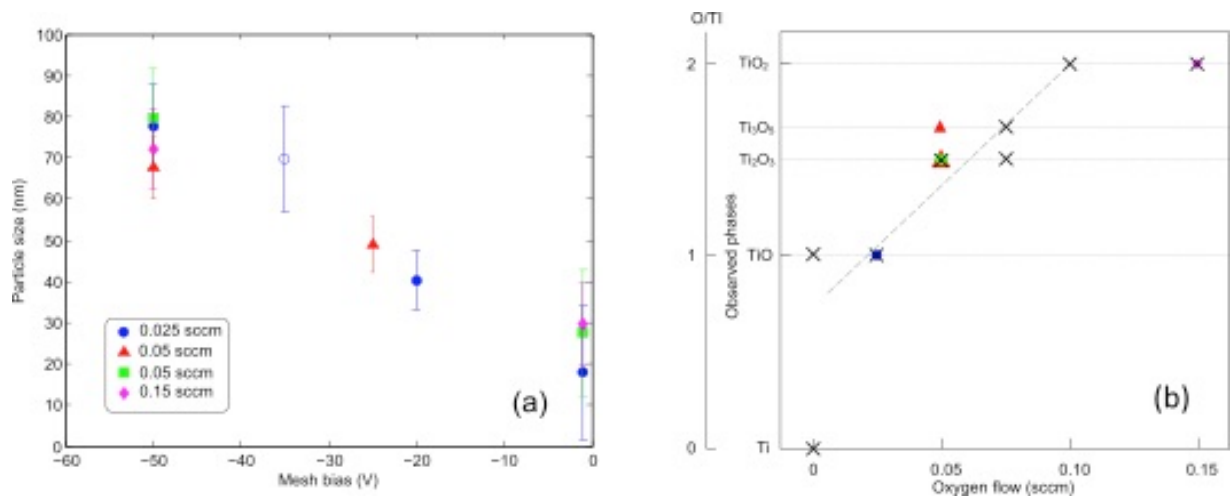

Fig. 6. The mean diameter with the standard deviation (a) and the phases observed with XRD (b) as functions of mesh bias and $O_2$ flows, respectively. Similar symbols in the figures indicate samples from the same series. The unfilled marker (-35 V) mean that no XRD was performed and the × symbol denotes NPs from other series than the one used in fig. a.

## Discussion

It is demonstrated that Ti-O NPs can reproducibly be synthesized with independent control of size and stoichiometry in a reactive gas atmosphere using pulsed hollow cathode sputtering. A main concern was that poisoning of the cathode could result in a hysteresis in the process, which would make the discharge unstable and the synthesis of NPs with desired composition difficult. However, the results demonstrate that the process is not in an unstable range and that the composition of the NPs can be tuned to any desired value from metallic with contaminants from the background pressure of the system, to fully oxidized $TiO_2$. The variation of the cathode voltage $U_D$ with the oxygen flow can be used to show that the discharge is operated below full poisoning of the cathode: the increase of $U_D$ at oxygen flows above 0.075 sccm (see Fig. 2, filled red squares) is understood as due an oxidation of the cathode surface by diffusion of $O_2$ to it. This increase continues above the $O_2$ flows needed to achieve fully stoichiometric $TiO_2$ nanoparticles, showing that the operation range in this work is characterized by only a partial cathode oxidation. From a process point of view, the avoidance of the poisoned mode can be beneficial for achieving a high productivity, since the sputter yield is higher in the metal mode than in the poisoned mode [4]. Such an advantage has been previously shown using a similar hollow cathode setup for thin film deposition of $Fe_2O_3$ [18].

The avoidance of cathode poisoning and pronounced hysteresis can be explained by one or more of the following three reasons: firstly, by using a hollow cathode design, where inert gas is flown through it and the reactive gas is separately introduced into the system, diffusion of the reactive gas to the inside of the cathode surface is inhibited. This limits poisoning and hinders the process from drifting to the unstable transition regime, as described for thin film processing using hollow cathodes [19]. Secondly, for a small enough cathode area where the sputter removal of oxidized layers is sufficiently efficient, a hysteresis-free operation of the reactive process is possible as has been demonstrated for thin film deposition [5]. Thirdly, high power pulsing has been experimentally shown to reduce hysteresis in high power impulse magnetron sputtering [6], with pulse lengths and amplitudes similar to the ones used here. To determine the relative importance of these three mechanisms, a more careful analysis and modeling needs to be done which is outside of the scope of the present work.

A small hysteresis that was observed could be related to the long time needed to reach steady state after altering the reactive gas flow. When the waiting time was longer the hysteresis decreased which was verified at 0.100 sccm both for increasing and decreasing flows. It is important to note that this is not the type of hysteresis often found in reactive sputtering, where the discharge can become unstable and jump between modes with different degrees of poisoning. No such unstable behavior was observed in the present discharge. The hystereses in Fig. 2 is instead due to the fact that the used waiting time is much shorter than the time it takes the discharge to establish equilibrium after each change in $O_2$ flow. This delay is probably characterized by the time needed to establish new oxidation levels at the walls and/or the cathode.

Process reproducibility is often a problem in reactive sputtering [4], and we find some also here when the oxygen flow is turned off (0 sscm) and for flows between 0.050 and 0.075 sccm. At 0 sccm the oxygen in the particles are most likely supplied from the partial pressure of the residual gas in the vacuum system. Since this partial pressure not only varies between experiments, but also during experiments (due to gettering), a variation in the particles should also be expected. For oxygen flows between 0.050 and 0.075 sccm, the synthesized phases $Ti_2O_3$ and $Ti_3O_5$ can both occur, and the crystalline phases measured at a constant flow could vary from experiment to experiment. These changes can also be caused by small variations in the residual gas pressure or by small changes in procedures during experiments. The experiment would be sensitive to such variations due to the long time to reach steady state.

The second main result presented in this work is the demonstration that the size of the NPs can be varied by altering a negative voltage that is applied to a mesh that surrounds the growth region. The exact mechanisms for this is not fully understood yet, but one important aspect is that the mesh can act as an independent second anode for the discharge, besides the grounded anode ring, and thereby deflect current away from the region in which the NPs grow. This influences the electron temperature and thereby the negative floating potential of the NPs. Their potential, in turn, determines the rate of trapping of positive metal ions from the plasma, and hence the NP growth rate. This connection between the NP growth rate and the electron temperature has previously been demonstrated by Pilch *et al* [16].

As demonstrated in the present work, the stoichiometry and the size of the particles can be varied independently. This means that there is no, or very small, crossover between parameters controlling the stoichiometry and the size, respectively. The small effect of size on stoichiometry is not surprising. Direct measurements show that a change in size (achieved through the mesh bias) of the NPs does not significantly affect the partial pressure of $O_2$. It is less obvious why a variation in stoichiometry (achieved through a variation of the $O_2$ flow) does not influence the size. On one hand: if the cathode were not poisoned at all, then the density of metallic growth material that is ejected into the growth zone would be unchanged. The NPs would then – to a first approximation – collect Ti at an unchanged rate, and in addition pick up oxygen. NPs grown at high oxygen flow, with a high content of oxygen, would then become larger. On the other hand: if the cathode was very strongly poisoned, then the sputtering of titanium would be strongly reduced and the NPs grow slower and become smaller. I addition to these opposing mechanisms that concern the NP growth rate, reactive species such as oxygen and nitrogen can have a positive influence on the nucleation of NPs [20]. Hence, an increase in the oxygen partial pressure might increase the rate of nucleation. Also here it is difficult to predict the effect on NP size. On one hand, a faster nucleation (sooner after the exit from the hollow cathode) would give individual NPs time to grow to larger size. On the other hand, the creation of a large density of nuclei might produce smaller nanoparticles, but at higher number density, as the source material is faster depleted.

We can only conclude that such opposing expected trends seem to balance each other in the preset device, and that both the nucleation process and the degree of poisoning of the cathode are likely to play important roles.

## Summary and Conclusions

Crystalline NPs of Ti-O with independently tunable size and oxygen content have been synthesized. By flowing Ar through the hollow cathode, a full oxidation of the cathode surface can be prevented which allowed to operate the process in the transition region between the metal and the poisoned mode. Fully stoichiometric $TiO_2$ NPs could be synthesized without fully poisoning the cathode, which results in higher deposition rate compared to sputtering from a poisoned cathode. A particular advantage of the presented setup is that it allows synthesizing NPs where the stoichiometry and the size can be controlled via two independent parameters, the reactive gas flow and the bias of a mesh surrounding the growth region. This mesh acts as a second anode to the discharge and can draw up to ¾ of the discharge current. The size of the NPs and its dependence on the mesh bias was well reproducible with only small variations. The stoichiometry was also well controllable with the exception that the crystalline phases in the intermediate region, $Ti_2O_3$ and $Ti_3O_5$, were less reproducible than TiO and $TiO_2$.

## List of abbreviations

Nanoparticles (NPs), residual gas analyzer (RGA), X-ray diffraction (XRD), scanning electron microscope (SEM), energy dispersive X-ray (EDX)

## Competing interests

Ulf Helmersson is co-fonder of TiÅ AB, a company with aim to exploit research results on NP-synthesis such as this study. The other authors declare no competing financial interests.

## Authors' contributions

RG carried out the experimental work and drafted the manuscript. IP supervised the work and IP and UH participated in the discussion of results and revised the manuscript. All authors read and approved the final manuscript.

## Authors' information

RG is a PhD student in the Plasma & Coatings Division at the Department of Physics, Chemistry and Biology at Linköping University, Sweden. He has been working with the synthesis of nanoparticles by pulsed hollow cathodes since he was an undergraduate student and has much experience with this technique. UH is professor in thin film physics and the head of the Plasma & Coatings Division at the Department of Division of Physics, Chemistry and Biology at Linköping University, Sweden. His research interests include synthesis of thin films and nanoparticles using advanced plasma techniques. IP received her Ph.D degree from Christian-Albrechts-Universität zu Kiel, Germany, in 2010. At present she is a research fellow in the Thin Film Division at the Department of Physics, Biology and Chemistry (IFM) at Linköping University, Sweden. Her research interests include plasma physics, complex plasmas, nanoparticle synthesis and diagnostic of nanoparticles during growth.

## Acknowledgment

This work has been financially supported by the Knut and Alice Wallenberg foundation (KAW 2014.0276) and the Swedish Research Council under Grant No. 2008-6572 via the Linköping Linneaus Environment LiLi-NFM. We like to thank Petter Larsson for designing key electronics used in this work and Daniel Magnfält for helping with the XRD characterization and discussions. We also would like to thank Joseph E. Greene, Ivan Petrov, and Nils Brenning for important suggestions.